# Crafting the strain state in epitaxial thin films: A case study of $CoFe_2O_4$ films on $Pb(Mg,Nb)O_3$-$PbTiO_3$

Zhiguang Wang,[1,*] Yaojin Wang,[1] Haosu Luo,[2] Jiefang Li,[1] and D. Viehland[1]

[1]Department of Materials Science and Engineering, Virginia Tech, Blacksburg, VA 24061, USA
[2]State Key Laboratory of High Performance Ceramics and Superfine Microstructure, Shanghai Institute of Ceramics, Chinese Academy of Sciences, Shanghai 201800, China

* To whom correspondence should be addressed. Email: zgwang@vt.edu

## Abstract:

The strain dependence of electric and magnetic properties has been widely investigated, both from a fundamental science perspective and an applications point of view. Electromechanical coupling through field-induced polarization rotation (PRO) and polarization reorientation (PRE) in piezoelectric single crystals can provide an effective strain in film/substrate epitaxial heterostructures. However, the specific pathway of PRO and PRE is a complex thermodynamic process, depending on chemical composition, temperature, electric field and mechanical load. Here, systematic studies of the temperature-dependent field-induced phase transitions in $Pb(Mg,Nb)O_3$-$PbTiO_3$ single crystals with different initial phase and orientation configurations have been performed. Different types of strains, volatile/nonvolatile and biaxial/uniaxial, have been measured by both macroscopic and in-situ X-ray diffraction techniques. In addition, the strain state of epitaxial Mn-doped $CoFe_2O_4$ thin films was examined by magnetic anisotropy measurements, where a giant magnetoelectric coupling has been demonstrated.

**KEYWORDS:** Magnetoelectric, magnetic anisotropy, phase transformation, electromechanical strain.

# I. INTRODUCTION

It is well known that both electric and magnetic properties of crystalline materials depend on their lattice parameters and the symmetry of their unit cell.[1-3] Strain engineering has been adopted to modulate the properties of a vast number of functional materials, including strained oxides, strained atomic sheets, strained catalysts, and strained silicon technologies.[4]

Recent advances in the growth of epitaxial thin films on single crystal substrates provide additional freedom in controlling the strain condition in thin films.[5-7] A layer-on-layer growth induced strong chemical bonding guarantees a near perfect strain transfer from substrate to film, whether an intrinsic lattice mismatch or an electrically controlled lattice deformation. With regards to strain engineering of ferroelectric thin films, the Curie temperature can be shifted hundreds of degree and the saturation polarization can be increased by more than a factor of two.[2,8,9] In ferromagnetic films grown on piezoelectric substrates, the magnetic easy axis can be rotated and the ferromagnetic resonance frequency can be shifted by the electrically induced strain of the substrate.[6,10,11] In superconductor films, the superconducting transition temperature can be shifted by strain.[12,13] In strongly correlated materials, the electrical conductivity can be changed by several orders due to strain from the substrate.[1]

Piezoelectric single crystals have been widely used as substrates due to a good crystal lattice parameter matching with most functional oxides and due to their large electromechanical coupling coefficients.[5,14,15] Significant effort has been placed on the study of electric field-induced strains in Pb-based piezoelectric crystals due to their applications in sensors and transducers. Most of these investigations have focused on strain evolution along the direction of the electric field, whereas the strain along the in-plane directions is the key factor for strain modulation of epitaxial films. One might anticipate that the in-plane strain can be estimated from the out-of-plane one based on Poission's ratio. However, detailed studies of the orientation-dependent mechanical properties of perovskites have shown an abnormal Poisson's ratio and compressibility that depend on both orientation and unit cell shape.[6,16] Moreover, previous studies treated the piezoelectric substrate as a linear strain source under application of external electric field, which is only true within a small range of electric fields over which phase transformations do not occur.

Here, Pb(Mg,Nb)$O_3$-PbTi$O_3$ (PMN-PT) substrates were selected with different initial phases (rhombohedral or tetragonal: R or T for short) and different out-of-plane orientations (<001>$_c$ or <011>$_c$, where "c" indicates pseudocubic notation). The temperature and electric field dependences of the phase transformations was thoroughly investigated by both macroscopic strain and microscopic in-situ XRD techniques. The strain effect was examined by measurement of the magnetic anisotropy of the Mn-doped CoFe$_2$O$_4$ (MCFO) thin films epitaxially grown on the substrates.[17]

## II. EXPERIMENT

Three different single crystals were used in this study: <001>$_c$-oriented PMN-30PT, <001>$_c$-oriented PMN-40PT and <011>$_c$-oriented PMN-29.5PT. To examine the strain effect on epitaxial films, MCFO thin films with thickness ~ 400 nm were deposited on PMN-PT substrates by pulsed laser deposition. Detailed growth process can be found in a previous report.[6] The crystal structure of the single crystal substrates and films was determined using a Philips X'pert high resolution X-ray diffractometer. The macroscopic strain curves were measured with capacitive displacement sensor under different electric fields. Magnetic hysteresis (M-H) loops were measured with a Lakeshore 7300 series vibrating sample magnetometer (VSM) system at room temperature, where the samples were connected to a Bertan high voltage power supply (210-20R) in order to apply electric fields. The field-induced strain from the piezoelectric PMN-PT substrate is transferred to the epitaxial MCFO thin film, thus changing the magnetic anisotropy distribution in the MCFO layer due to the well-known magnetoelectric effect of the heterostructure.[14,15,17-23]

## III. RESULTS AND DISCUSSION

Figure 1 shows the three factors that determine the type of strain that can be obtained from PMN-PT substrates. First, the initial phase configuration of the PMN-PT substrate at room temperature is defined by the composition of $(1-x)$PMN-$x$PT. PMN-PT with low PT contents ($x<0.3$) has a rhombohedral (R) structure with $P_s$ oriented along <111> directions. A mixture of R, orthorhombic (O), monoclinic (M) and tetragonal (T) phases is found for $0.3<x<0.38$.[24] For $x>0.38$, the stable phase is tetragonal. Application of electric field can change the polarization direction, distorting the unit cell to a different symmetry. This in turn results in a strain in the epitaxial film. Field-induced polarization rotation depends on both the direction along which the electric field is applied and the ambient temperature conditions.[25]

Figure 2 shows the electric field dependent polarization rotation pathways and resultant strain conditions for PMN-PT of several different orientations and different initial phases. The R-phase has eight equivalent polarization directions (Fig.2a). For $E//<001>_c$, the polarization will rotate away from $<111>_c$ towards $E$ within monoclinic A ($M_C$) and then C ($M_C$) phases. At sufficiently high $E$, an induced R→T transition occurs,[26] resulting in a pseudo uniform strain in the heteroepitaxial film. The polarization as well as the long axis direction of the field-induced T phase is along out-of-plane, therefore, the field-induced in-plane strain is compressive. In the T-phase, there are six equivalent polarization directions (Fig.2b). The ferroelectric domains can be divided into two types: $c$-domains whose polarization lies along out-of-plane and $a$-domains whose polarization is in-plane. By application of $E//<001>_c$, $a$-domains will transform to $c$-domains, inducing a large strain in the film on the crystal substrate.[27] This type of strain is strongly related to the original local domain state: for $a$-domain regions, the $a$ to $c$-domain transition results in a large strain, whereas the original $c$-domain regions exhibit much smaller strain. For <011>$_c$-oriented R-phase crystals, there are eight equivalent polarization directions (Fig.2c). Under $E$//<011>$_c$, the polarization will rotate away from the <111>$_c$ in a M$_B$ phase towards <011>$_c$. Under a sufficiently high $E$, a mono-domain O-phase will be induced.[6] During this R→M$_B$→O phase transition, a uniaxial strain is generated in the heteroepitaxial film, which is compressive along IP <100>$_c$ and tensile along IP <0-11>$_c$ (Fig.2c).

Figure 3 shows macroscopic unipolar strain curves for <001>$_c$ PMN-30PT crystals at different temperatures. A phase transformation sequence of R→M$_A$→M$_C$→T has been widely reported in PMN-PT for $E$//<001>$_c$.[28,29] During $P$ rotation in the monoclinic phases, the OP component of $P$ continuously increases whereas the IP one decreases. Thus, a uniform compressive strain is induced in an heteroepitaxial thin film deposited on such a PMN-PT substrate. Three different stages of the strain process can be identified, indicating different phase transformation steps (Fig.3a). The shape changes of the R→M$_A$→M$_C$ phase transformation sequence is tempered due to the relatively flat free energy profile across these phases,[28] however the M$_C$→T phase transformation induces a significant distortion of the unit cell and consequently a sharp increase of $\varepsilon$. We denote the field at which the M$_C$→T transformation occurs with a large change in $\varepsilon$ as $E_{M-T}$. Figure 3(b) summarizes the relationship between $E_{M-T}$ and temperature. Since the T-phase is the higher temperature phase, the M$_C$→T transformation becomes easier with increasing temperature, and thus $E_{M-T}$ is decreased. When used as a substrate, this shows that a much larger strain can be transferred to a heteroepitaxial layer by relatively smaller fields by proper selection of temperature. Figure 3(c) shows XRD line scans under different $E$ where a left shift can clearly be identified due to the elongation of the crystal lattice parameter with increasing $E$. The in-situ equivalent out-of-plane crystal lattice parameter ($c_{op1}$) as a function of $E$ is shown in Fig.3(d). The value of $c_{op1}$ was calculated by the formula

$$c_{op1} = 2d = \frac{\lambda}{2\sin(\theta)}, \quad (1)$$

where λ is the wavelength of the X-ray source with value λ=1.5406 Å. For as-grown PMN-30PT, $c_{op1}$=1.5604/sin(45.02°/2)=4.024 Å, which subsequently increased in value with increasing $E$. The first jump near $E$=2 kV/cm represents the R→$M_A$ transformation. After entering the monoclinic phase region, the $\varepsilon$-$E$ response was a near linear curve as long as $E$ was not sufficient to induce a $M_C$→T transition near $E$≈30kV/cm. After removal of $E$, the $M_A$ phase configuration is maintained, resulting in an irreversible increase of $c_{op1}$. The strain can be calculated by the equation

$$\varepsilon = \frac{\Delta c_{op1}}{c_{op1}} = \frac{c_{op1}@10kV/cm - c_{op1}@0kV/cm}{c_{op1}@0kV/cm}, \qquad (2)$$

Accordingly, on the first cycle of the unipolar $\varepsilon$-$E$ curve, the field-induced strain reached a value of $\varepsilon$ =0.32% whereas on the second cycle, it reached only $\varepsilon$=0.2%. This difference is due to the irreversible R→$M_A$ transformation. The first time field-dependent in-plane strain is complex due to the complex R→$M_A$→$M_C$ phase transition sequence. An reversible in-plane strain $\varepsilon_{ip}$=0.028% @10kV/cm can be obtained from the microscopic XRD measurements, which can then subsequently be transferred to a heteroepitaxial magnetic layer for the modulation of the magnetic anisotropy.

Figure 4 shows the electric field dependence of the magnetic hysteresis (M-H) loops along both IP and OP. The M-H loop change as a function of electric field is hardly recognizable due to the small in-plane strain. MCFO has a large negative magnetostriction. Therefore, an compressive strain decreases its magnetic anisotropy along IP, whereas a tensile one increases it along OP. Figure 4(c) shows the magnetization squareness factor $R_M$=$M_r$/$M_s$ as a function of $E$. Because the IP strain is very small, the strain induced magnetic anisotropy change is much smaller. Values of $\Delta R_M$ = 0.01 and 0.002 (0→10kV/cm) were found along IP and OP directions, respectively.

Figure 5 shows the details of $E$-dependent $P$ reorientation in <001>$_c$-oriented PMN-40PT which has a stable T-phase ($a$=$b$<$c$ and $\alpha$=$\beta$=$\gamma$=90°). As-grown PMN-40PT has six possible equivalent polarization directions, as shown in Fig.5(a). We denote the domains with $P$ along IP as $a$- domains, whereas these along OP as *c*-domains. For the *a*-domains, the two in-plane crystal lattice parameters are *a* and *c*, whereas the out-of-plane one is *a*. For the *c*-domains, both in-plane lattice parameters are *a*, whereas the out-of-plane one is *c*. After *a*→*c* domain reorientation, a large in-plane compressive strain was induced. Figure 5(b) shows XRD measurements of the *E* dependent domain configuration (ratio of *a* and *c* domains) in PMN-40PT. The measurement was performed along OP direction, thus the $(002)_c$ peak with a larger d-spacing parameter comes from *c*-domains, whereas $(200)_c$ comes from *a*-domains. The tetragonal crystal lattice parameters were calculated to be (*a*, *c*)=(3.996Å, 4.051Å). Thus, for the completed *a*→*c* domain rotation, a compressive strain of $\varepsilon=(a-c)/c\approx$ -1.36% is obtained along IP. The total IP compressive strain value is the statistical average value over all the domains. The OP tensile strain can then be estimated by the *a*-domain/*c*-domain ratio before and after application of *E* based on the intensity of the diffraction peaks from each domain component, as shown in Fig.5(c). A tensile OP strain of $\varepsilon=1.2\%$ and a compressive IP strain of $\varepsilon=$ -0.6% can thus be obtained.

Figure 6 shows the electric field dependence of the M-H loops along both the IP and OP directions of a MCFO/<001>$_c$PMN-40PT heterostructure. Compressive strain along IP increases the magnetic anisotropy along that direction, whereas tensile strain decrease it along OP. Accordingly, the remnant magnetization along IP is decreased, while that along OP is increased. Figure 6(c) shows the $R_M$ value as a function of *E*. The values of $R_M$ along both IP and OP are stable for 0<*E*<10 kV/cm due to a stable domain configuration. However, as *E* is further increased, the *a*-domains begin to be switched to *c*-ones, inducing a large IP compressive strain. In turn, this results in a sharp increase of $R_M$ value along IP and a sharp decrease of that along OP for 11<*E*<14 kV/cm. For *E*> 14 kV/cm, the *a*→*c* domain transition is completed, thus the strain state as well as the $R_M$ values become stable. Upon removal of *E*, the field-induced *a*→*c* domain configuration is stable, and thus the strain condition is maintained. This is evidenced by both the XRD measurement and stable $R_M$ values with decreasing *E* from 16 to 0 kV/cm, as shown in Fig.6(c). Although it is an irreversible one-time effect, the strain induced $\Delta R_M$ for the MCFO film on PMN-40PT is more than ten times larger than that of the MCFO film on $<001>_c$ PMN-30PT.

Figure 7 shows the $E$ dependence of the $P$ rotation in $<011>_c$-oriented PMN-29.5PT. Initially the crystal had a stable R-phase with eight possible equivalent polarization directions, as shown in Fig.7(a). Under an out-of-plane $E$, a transformation sequence of R→$M_B$→O has been widely reported.[30-33] The application of $E$ degenerates the eight domain states into the two acclivous ones, as shown in Fig.7(b). With further increase of $E$, the two $M_B$ domain states transform to a single domain state that has a stable mono-domain O-phase, as shown in Fig.7(d). Upon removal of $E$, the stable phase relaxes back to $M_B$ with two domain states. Please note that this relaxation back to a polydomain M phase is strongly composition dependent;[6] and with slight increase of PT content, the O phase can be retained upon removal of $E$.

Figure 8(a) shows XRD line scans for the $E$ dependent domain configuration of PMN-29.5PT. Measurements were performed along OP, and the R-phase (022) peak indicates a inter-planar spacing of $d_1$=2.845 Å. Under $E$=10kV/cm, a mono-domain O-phase condition was induced with an inter-planar spacing of $d_2$=2.851 Å. The out-of-plane strain was determined to be $\varepsilon_{op}=\Delta d/d=(2.851-2.845)/2.845=0.21\%$ based on the changes in the (022) inter-planar spacing, which was further confirmed to be $\varepsilon_{op}$=0.26% based upon the macroscopic strain measurement, as shown in Figs.8(b) and (c). Figure 8(d) shows XRD line scans under different $E$. For $E$=0 kV/cm, a diffraction peak of R-(200)$_c$ (2θ=42.02°) was observed, which then shifted with increasing $E$ (0→5 kV/cm) to O-(200)$_c$ (2θ=42.156°). Upon removal of $E$, the diffraction peak shifted back to the original position due to the relaxation of the field-induced O-phase. Therefore, a reversible IP strain of $\varepsilon_{ip}=\Delta d/d$=0.56% which is two times of $\varepsilon_{op}$, can be obtained with $E$=5 kV/cm. More importantly, this is a transverse strain that can be effectively transferred to the epitaxial MCFO thin films, and thus can result in dramatic changes in the magnetic anisotropy distribution.

Figure 9 shows the electric field dependence of the M-H loops taken along both in-plane and out-of-plane directions for a MCFO/<011>$_c$-PMN-29.5PT heterostructure. One should note that the field-induced strains along IP<0-11>$_c$ and OP<011>$_c$ are tensile whereas that along IP<100>$_c$ is compressive. Therefore, the tensile strains will decrease the magnetic anisotropy along IP<0-11>$_c$ and OP<011>$_c$, whereas the compressive strain will increase the magnetic anisotropy along IP<100>$_c$. Accordingly, the remnant magnetizations along IP<0-11>$_c$ and OP<011>$_c$ are decreased with increasing $E$, whereas that along IP<100> is increased, as shown in Fig.9 (a-c). Figure 9 (d) shows the magnetization squareness factor R$_M$ as a function of $E$. The strain modulation of the R$_M$ value happened simultaneously along all three directions characterized. A sharp change was observed near $E$=4 kV/cm, indicating a critical electric field for the M$_B$→O transformation. The strain value was relatively stable for $E$>5 kV/cm as shown in Fig.9 (b) and (c). Accordingly, the R$_M$ values along all three directions were stable for $E$>5 kV/cm. Moreover, hysteresis behaviors are obvious in both the $\varepsilon$-$E$ and R$_M$-$E$ curves, reflecting a hysteresis in the O→M$_b$ phase relaxation as $E$ is decreased.

Figures 10(a) and (b) summarize the field dependent strains for PMN-PT crystals of different compositions and orientations. The specific electric field on each sample was determined by the threshold $E$ value for polarization rotation/reorientation. A large IP strain of $\varepsilon\approx$ -0.68% was observed in <001>$_c$-oriented PMN-40PT under $E$=15 kV/cm. However, this was a one-time effect that could not be reversed by $E$. On the contrary, a reversible strain of $\varepsilon\approx$ -0.56% along IP <100>$_c$ was found for <011>$_c$-oriented PMN-29.5PT, which could easily be modulated by $E$. The reversible IP strain for <001>$_c$-oriented PMN-30PT was very small ($\varepsilon\approx$-0.02%).

Figure 10(c) summarizes the magnetization squareness modulation results for the heterostructures with different electric field histories. IP<100>$_c$ is selected here as the direction for comparisons. For MCFO film on <001>$_c$ PMN-30PT, the value of |ΔR$_M$| was stable but small (|ΔR$_M$|=0.01). During the first measurement cycle of MCFO on <001>$_c$ PMN-40PT and on <011>$_c$ PMN-

29.5PT, large field-induced strains resulted in large changes in the distribution of the magnetic anisotropy: $|\Delta R_M|$=0.128 and 0.268, respectively. However, during the second measurement cycle, the field-induced reversible strain in PMN-40PT was small, and thus $|\Delta R_M|$=0.0001, as shown in Fig.10(c). On the contrary, the reversible strain for PMN-29.5PT is high during the second cycle due to the relaxation of the field-induced O-phase. Accordingly, the value of $|\Delta R_M|$=0.258 during the second measurement cycle (Fig.10c) was just as high as that in the first cycle. Furthermore, this value was about 26 times that of MCFO on <001> PMN-30PT. The dramatic difference in the value of $|\Delta R_M|$ shows that changes of $|\Delta R_M|$ is not only related to the strain value along that specific direction, but also related to the symmetry of the interface strain that has been transferred from the substrate to the epitaxial thin film. The MCFO thin film has a large magnetic shape anisotropy along in-plane owning to the drastic dimension difference between length and thickness directions, and thus the magnetic moments have been confined within in-plane even before application of electric-field. Therefore the electric-field induced uniform IP magnetic strain anisotropy in MCFO on $<001>_c$ PMN-30PT has to compete with magnetic shape anisotropy to induce a change in the magnetization squareness along both IP and OP direction. For MCFO on $<011>_c$ PMN-29.5PT, the magnetic strain anisotropy decreases along IP<0-11> due to a tensile strain, whereas that increases along IP<100> due to a compressive strain. No magnetic shape anisotropy difference exists between the two in-plane directions. Therefore, the uniaxial in-plane strain in $<011>_c$ PMN-29.5PT is much more effective compared with a relatively uniform strain in $<001>_c$-oriented PMN-PT in the concern of magnetic property modulation.

In summary, we have systematically studied the field-induced strain in epitaxial MCFO films on PMN-PT single crystals with different initial phase stabilities and crystal orientations. MCFO thin films on $<001>_c$ PMN-30PT and $<011>_c$ PMN-29.8PT showed reversible strains due to relaxation of the field-induced phases. The change in magnetization squareness of MCFO on $<011>_c$ PMN-29.5PT is about 26 times of that on <001>$_c$ PMN-30PT, which is due to competing magnetic strain and shape anisotropies. Therefore, it is unambiguous that the uniaxial strain in <011>$_c$-oriented PMN-29.5PT is much more effective in controlling the direction of related properties, e.g. magnetization reversal, conductivity, electron mobility, etc. The strain in <001>$_c$ PMN-40PT is a one-time effect. These findings are generally applicable to all epitaxial films on piezoelectric single crystal substrates, where strain engineering might be used for exploration of new memory and logic devices based on multiferroic heterostructures.

## ACKNOWLEDGMENTS

This work was supported by U.S. Department of Energy (DE-FG02-06ER46290).

# Figure Captions

**Figure 1.** Schematic of the three factors that determine the field-induced strain amplitude and stability.

**Figure 2.** Schematic of the electric field induced polarization rotation and reorientation in PMN-PT single crystals and the resultant strains with different directivity: (a) <001>$_c$-oriented PMN-30PT; (b) <001>$_c$-oriented PMN-40PT; (c) <011>$_c$-oriented PMN-29.5PT.

**Figure 3.** (a) Field-induced OP strain at different temperature in the <001>$_c$-oriented PMN-30PT. (b) Temperature dependent threshold electric field for $M_c \rightarrow T$ phase transformation. (c) XRD characterization of the <002>$_c$ peaks shifted by external electric fields. (d) Electric field dependent OP crystal lattice parameters.

**Figure 4.** Field dependent M-H loops of MCFO film on <001>$_c$-oriented PMN-30PT substrates along IP (a) and OP (b) directions, respectively. (c) Field dependent magnetization squareness ($R_M$).

**Figure 5.** (a) Field-induced polarization reorientation in <001>$_c$-oriented PMN-40PT. (b) XRD results of PMN-40PT under different electric fields. (c) Electric field dependent OP d-spacing parameter.

**Figure 6.** Field dependent M-H loops of MCFO film on <001>$_c$-oriented PMN-40PT substrates along IP (a) and OP (b) directions, respectively. (c) Field dependent magnetization squareness ($R_M$).

**Figure 7.** Field-induced polarization rotation in <011>$_c$-oriented PMN-29.5PT.

**Figure 8.** (a) XRD results of <011>$_c$ PMN-29.5PT under different electric fields. (b) Electric field dependent OP d-spacing parameter. (c) Field dependent macroscopic strain curve. (d) XRD results along IP<100>$_c$ under different electric fields.

**Figure 9.** Field dependent M-H loops of MCFO film on <011>$_c$-oriented PMN-29.5PT along IP<0-11>$_c$ (a), IP<100>$_c$ (b), and OP<011>$_c$ (c) directions, respectively. (d) Field dependent R$_M$.

**Figure 10.** Field dependent strain amplitudes in PMN-PT single crystals with different composition and characterizing directions at first time (a) and after first time (b). (c) Field dependent magnetization squareness in MCFO films on different PMN-PT single crystals.

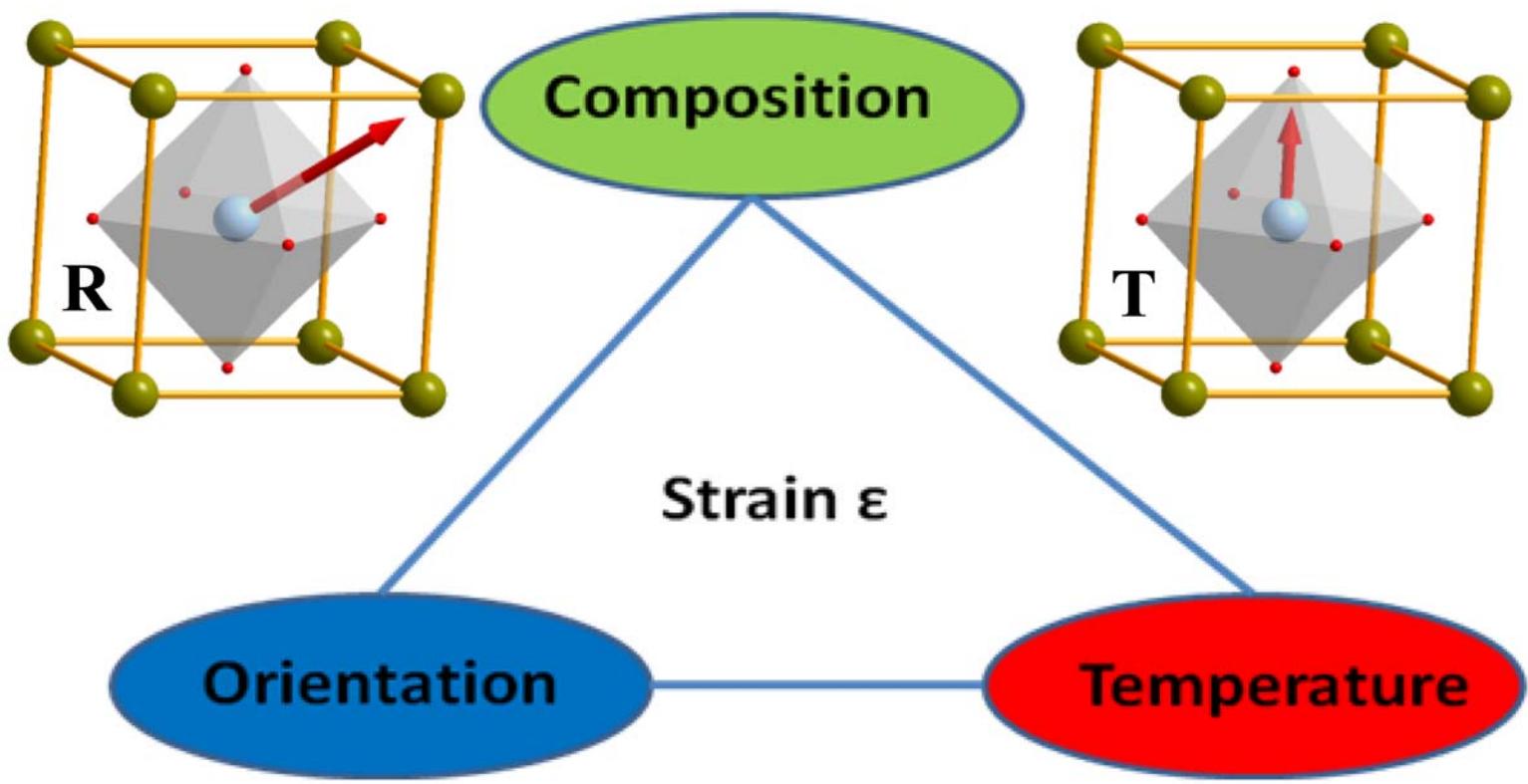

Fig.1

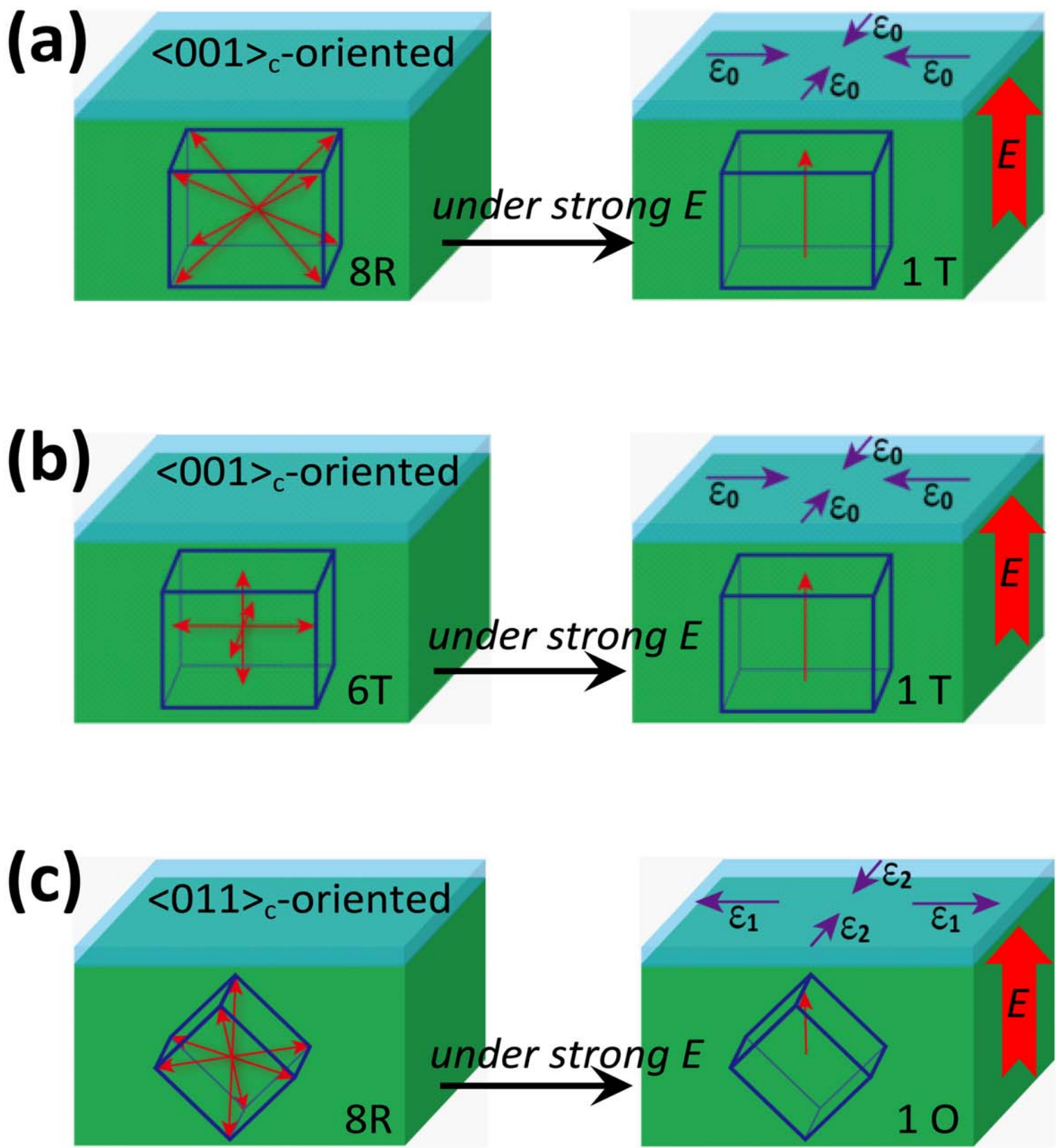

Fig.2

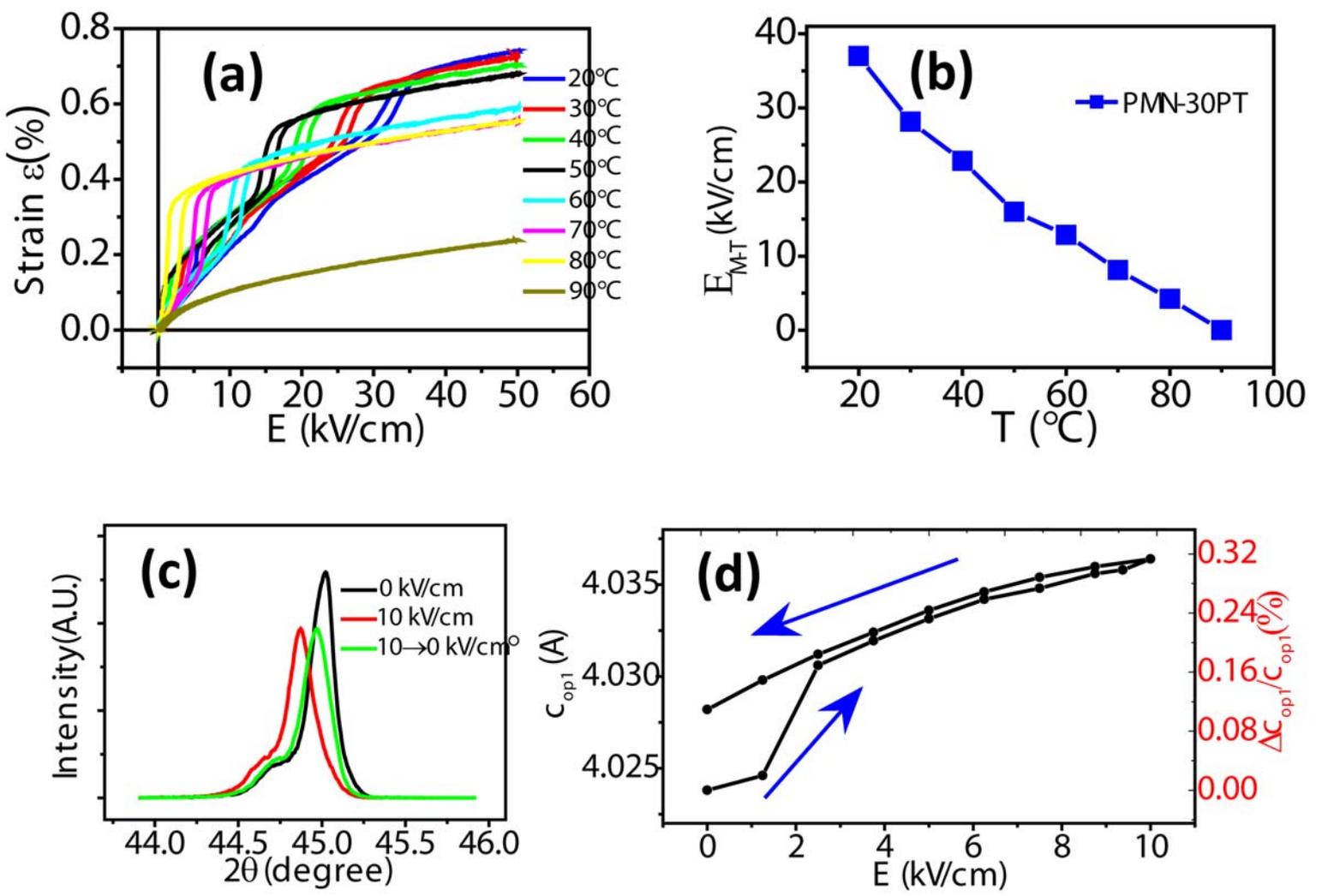

Fig.3

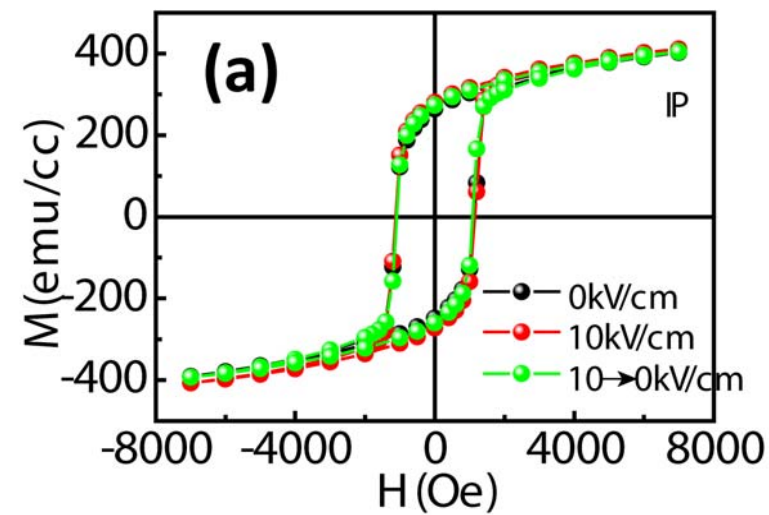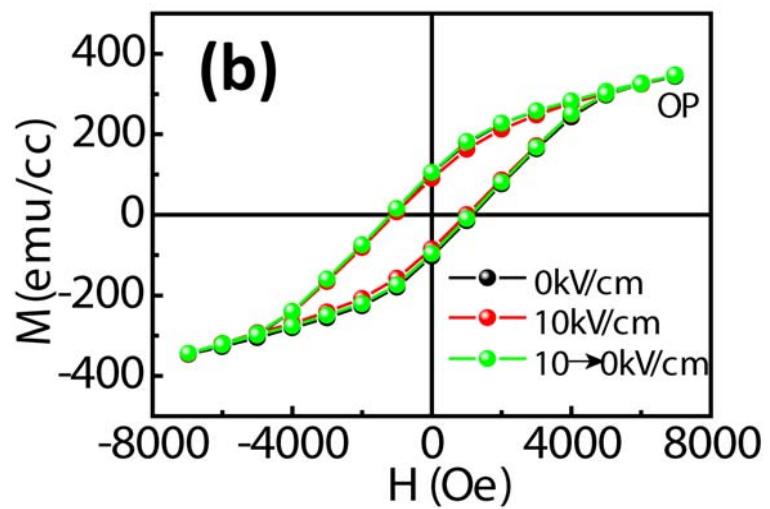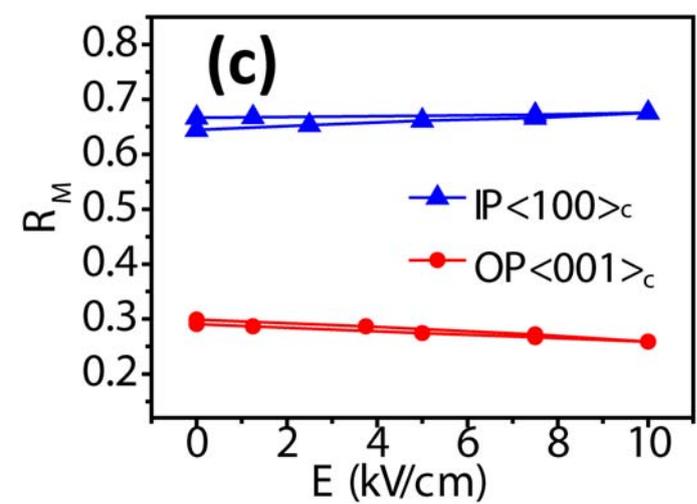

Fig.4

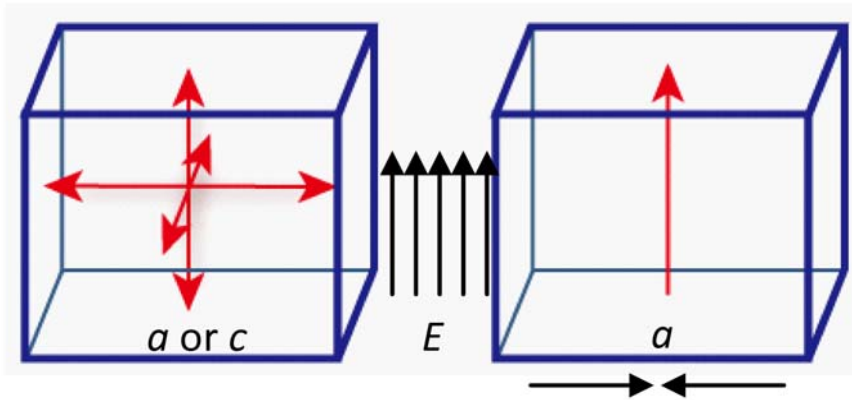
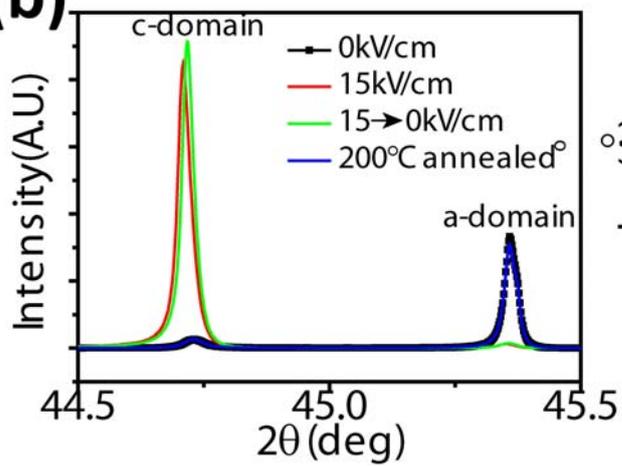
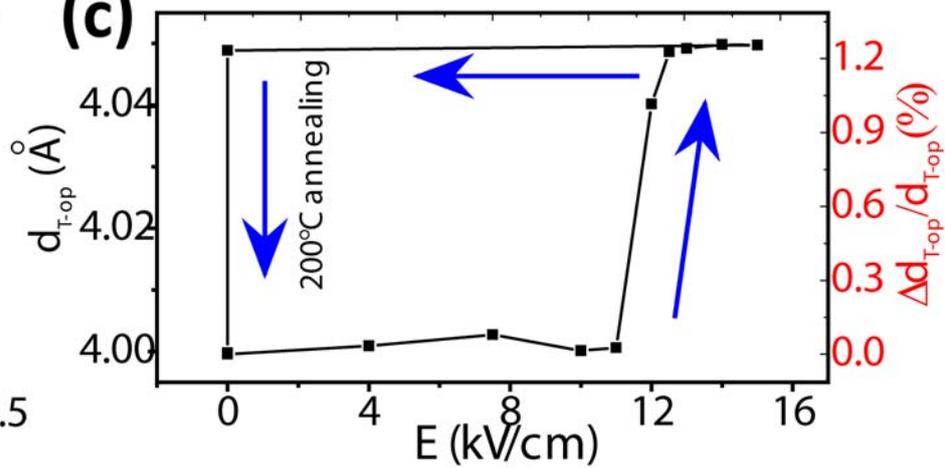

**Fig.5**

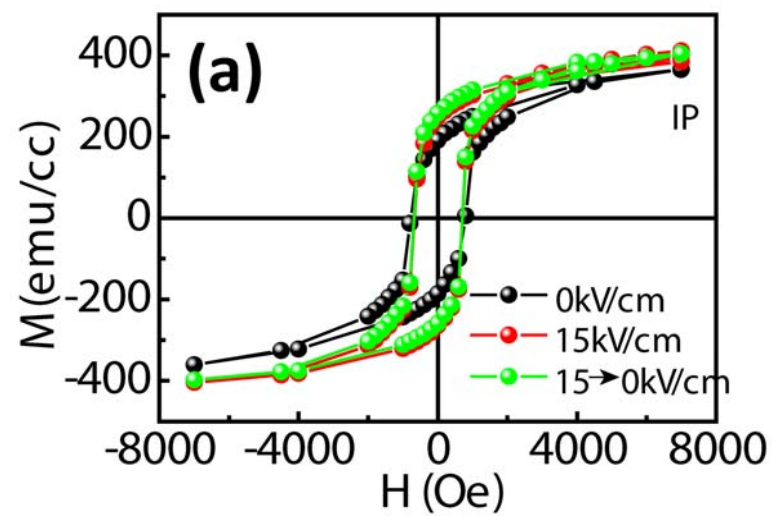
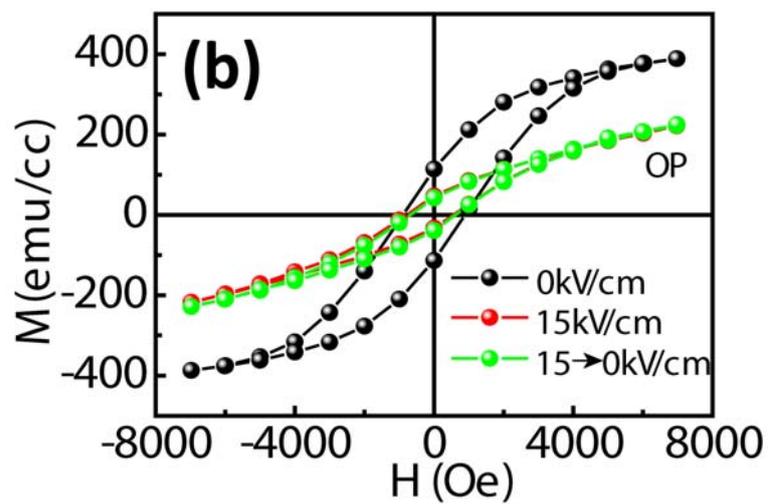
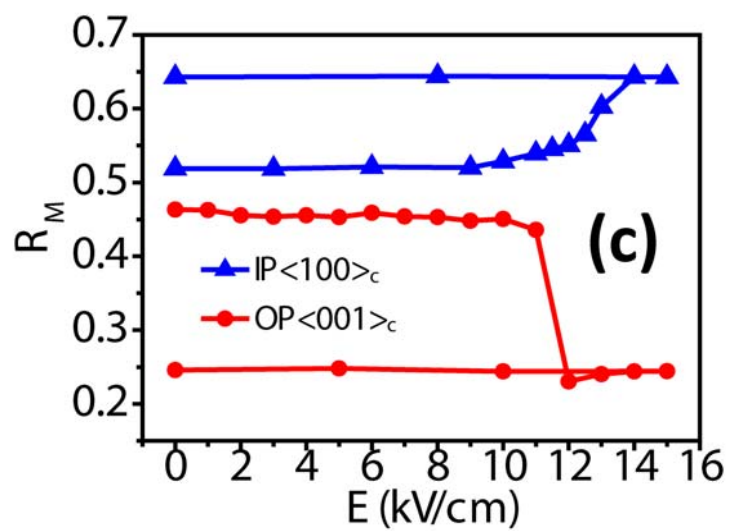

Fig.6

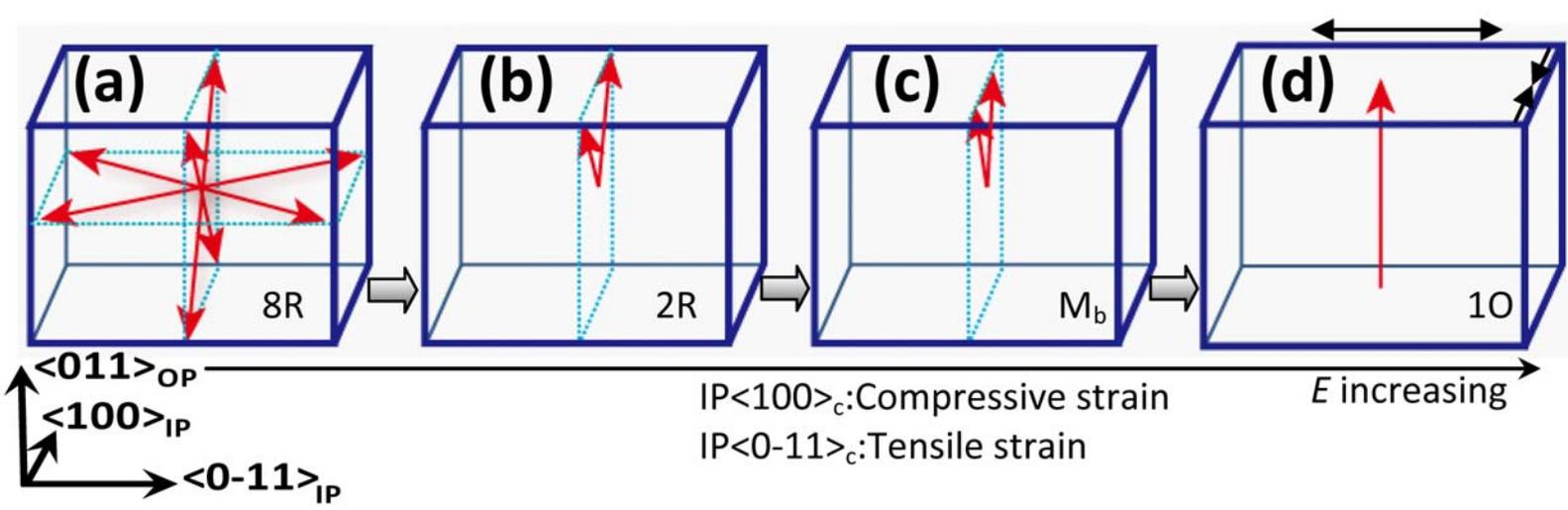

Fig.7

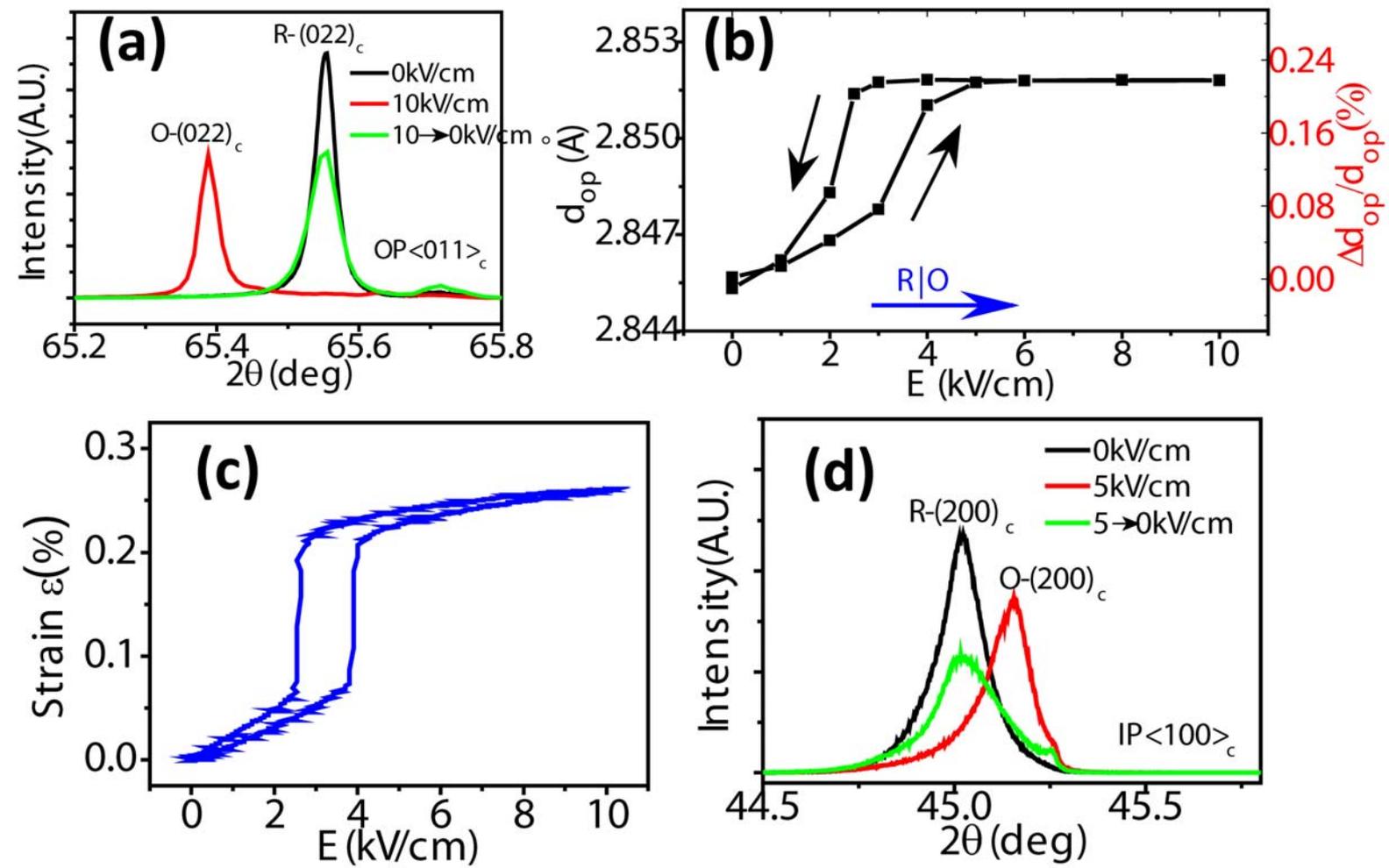

Fig.8

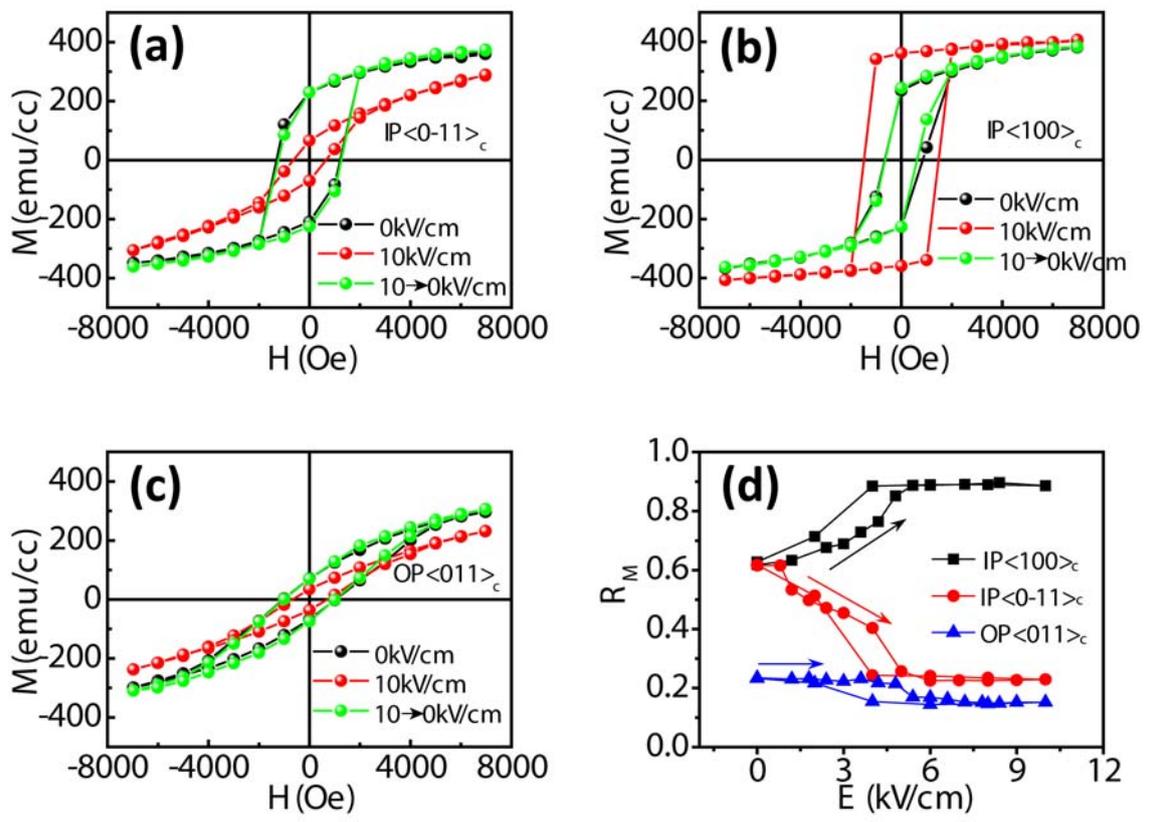

Fig.9

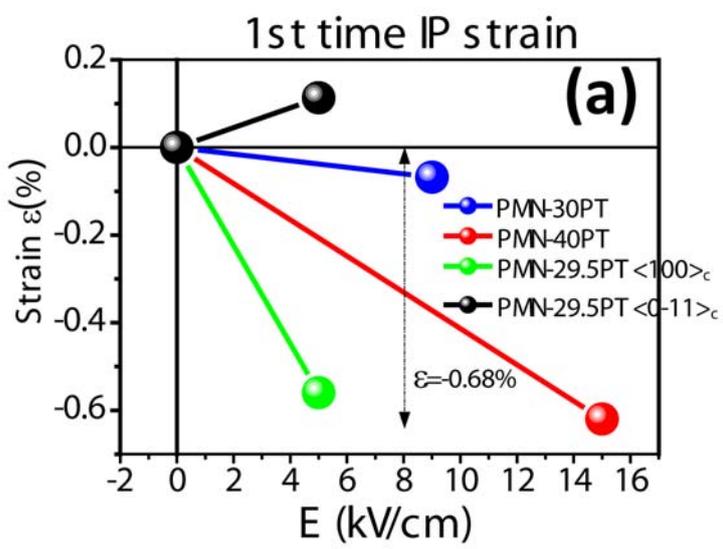
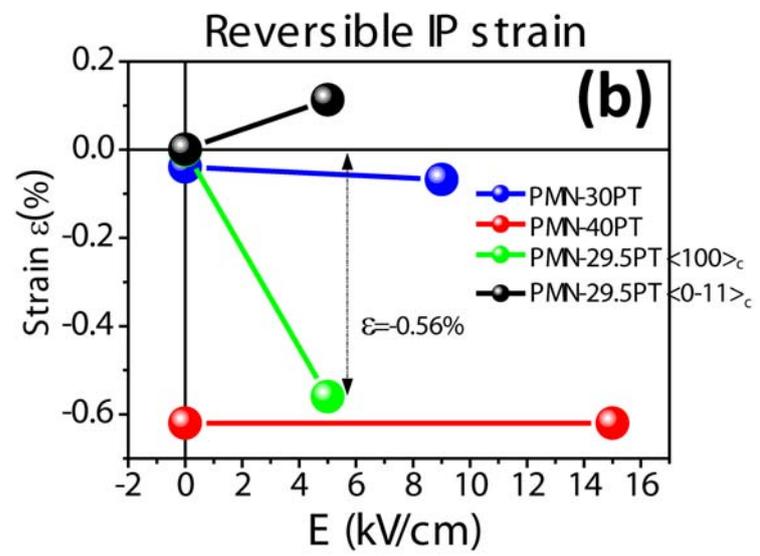
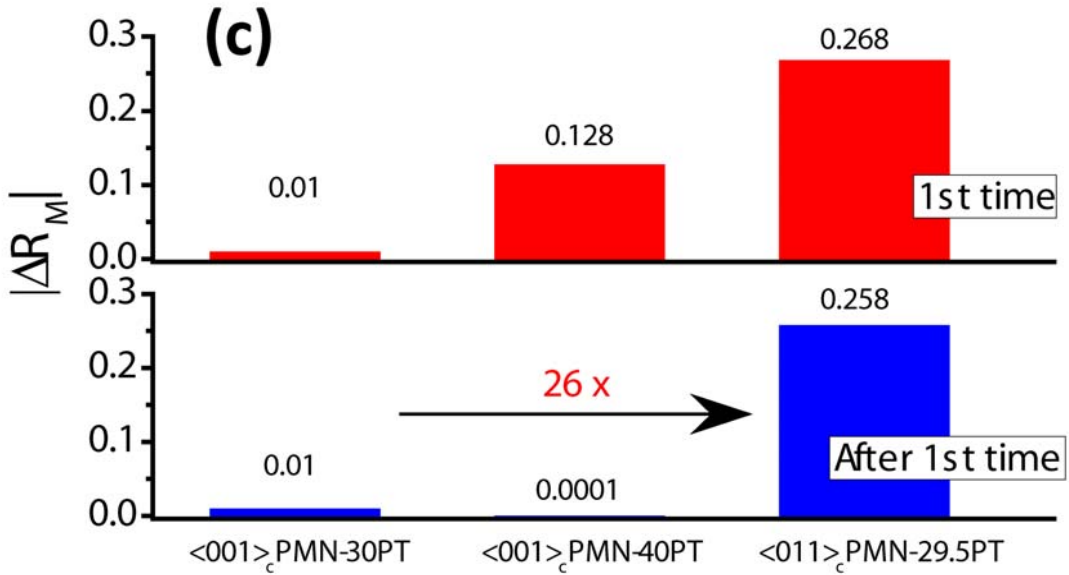

Fig.10